\documentclass[preprint2, tighten, usenatbib, appendixfloats]{aastex63}
\usepackage{graphicx} 

\newcommand{\sn}{SN~2024aihh}
\def\arcsec{\hbox{$^{\prime\prime}$}}

\begin{document}

\title{JWST and HST observations of the host galaxy and supernova, \sn~ in EP240801a at z=1.67}

\shortauthors{van Hoof et al.}
\newcommand{\NU}{\affiliation{Center for Interdisciplinary Exploration and Research in Astrophysics (CIERA) and Department of Physics and Astronomy, Northwestern University, Evanston, IL 60208, USA}}

\newcommand{\Radboud}{\affiliation{Department of Astrophysics/IMAPP, Radboud University, 6525 AJ Nijmegen, The Netherlands}}

\newcommand{\Leicester}{\affiliation{School of Physics and Astronomy, University of Leicester, University Road, Leicester, LE1 7RH, UK}}

\newcommand{\LANL}{\affiliation{Center for Nonlinear Studies, Los Alamos National Laboratory, Los Alamos, NM 87545 USA}}

\newcommand{\ESA}{\affiliation{European Space Agency (ESA), European Space Astronomy Centre (ESAC), Camino Bajo del Castillo s/n, 28692 Villanueva de la Cañada, Madrid, Spain}}

\newcommand{\INAFbologna}{\affiliation{INAF–Osservatorio di Astroﬁsica e Scienza dello Spazio, via Piero Gobetti 93/3, I-40129 Bologna, Italy}}

\newcommand{\INAFRoma}{\affiliation{INAF–Osservatorio Astronomico di Roma, via Frascati 33, I-00040 Monte Porzio Catone, Italy}}

\newcommand{\Caltech}{\affiliation{Department of Astronomy and Astrophysics, California Institute of Technology, Pasadena, CA 91125, USA}}

\newcommand{\Birmingham}{\affiliation{School of Physics and Astronomy, University of Birmingham, Birmingham B15 2TT, UK}}

\newcommand{\BirmIGWA}{\affiliation{Institute for Gravitational Wave Astronomy, University of Birmingham, Birmingham B15 2TT}}

\newcommand{\IfAUH}{\affiliation{Institute for Astronomy, University of Hawaii, 2680 Woodlawn Drive, Honolulu, HI 96822, USA}}

\newcommand{\Oxford}{\affiliation{Astrophysics sub-Department, Department of Physics, University of Oxford, Keble Road, Oxford, OX1 3RH, UK}}

\newcommand{\Belfast}{\affiliation{Astrophysics Research Centre, School of Mathematics and Physics, Queen’s University Belfast, BT7 1NN, UK}}
\author[0009-0005-5404-2745]{Agnes P.C. van Hoof}
\Radboud

\author[0000-0001-7821-9369]{Andrew J. Levan}
\Radboud
\affil{Department of Physics, University of Warwick, Coventry, CV4 7AL, UK}

\author[0000-0001-5679-0695]{Peter G. Jonker}
\Radboud

\author[0000-0003-2191-1674]{Morgan Fraser}
\affiliation{School of Physics and Centre for Space Research, University College Dublin, Belfield, Dublin 4, Ireland}

\author[0000-0001-9695-8472]{Luca Izzo}
\affil{Osservatorio Astronomico di Capodimonte, INAF, Salita Moiariello 16, Napoli, 80131, Italy}
\author{Andrew S. Fruchter}
\affil{Space Telescope Science Institute, 3700 San Martin Drive, Baltimore, MD21218, USA}

\author[0009-0007-6927-7496]{Joyce N. D. van Dalen}
\Radboud

\author[0000-0003-3274-6336]{Nial R. Tanvir}
\affiliation{School of Physics and Astronomy, University of Leicester, University Road, Leicester, LE1 7RH, UK}
\author[0000-0002-4571-2306]{Jens Hjorth}
\affil{DARK, Niels Bohr Institute, University of Copenhagen, Jagtvej 155A, 2200 Copenhagen, Denmark}

\author[0000-0001-5108-0627]{Antonio Martin-Carrillo}
\affiliation{School of Physics and Centre for Space Research, University College Dublin, Belfield, Dublin 4, Ireland}

\author[0000-0003-2700-1030]{Nikhil Sarin}
\affiliation{{Kavli Institute for Cosmology Cambridge}, {{Madingley Road}, {Cambridge CB3 0HA}, {United Kingdom}}}
\affiliation{{Institute of Astronomy}, {University of Cambridge}, {{Madingley Road}, {Cambridge CB3 0HA}, {United Kingdom}}}


\author[0000-0002-7910-6646]{Laura C. Cotter}
\affiliation{School of Physics and Centre for Space Research, University College Dublin, Belfield, Dublin 4, Ireland}

\author[0000-0001-8602-4641]{Jonathan A. Quirola Vasquez}
\Radboud

\author[0000-0003-3193-4714]{Maria E. Ravasio}
\Radboud

\author[0000-0003-2276-4231]{Javi S\'anchez-Sierras}
\Radboud

\author[0000-0002-5297-2683]{Manuel A. P. Torres}
\affil{Instituto de Astrof\'isica de Canarias, E-38205 La Laguna, Tenerife, Spain}
\affil{Departamento de Astrofísica, Univ. de La Laguna, E-38206 La Laguna, Tenerife, Spain}






\author[0009-0001-8155-7905]{Shuai-Qing Jiang}
\affiliation{National Astronomical Observatories, Chinese Academy of Sciences, Beijing 100101, People's Republic of China}
\affiliation{School of Astronomy and Space Science, University of Chinese Academy of Sciences, Chinese Academy of Sciences, Beijing 100049, People's Republic of China}

\author[0000-0003-3257-9435]{Dong Xu}
\affiliation{National Astronomical Observatories, Chinese Academy of Sciences, Beijing 100101, People's Republic of China}
\affiliation{Altay Astronomical Observatory, Altay, Xinjiang 836500, People's Republic of China}

\correspondingauthor{Agnes van Hoof}
\email{agnes.vanhoof@ru.nl}

\begin{abstract}
We present James Webb Space Telescope ({\em JWST}) and Hubble Space Telescope ({\em HST}) observations of the counterpart of EP~240801a, at $z=1.67$, the first fast X-ray transient (FXT) identified as an X-ray flash (XRF) by the Einstein Probe (EP) and {\em Fermi}-GBM.
Our observations reveal strong photometric and spectroscopic evidence for an associated broad-lined Type Ic supernova (SN) \sn, the most distant spectroscopically identified gamma ray-burst (GRB)--SN to date. The SN exhibits similar luminosity and light curve evolution to the proto-type GRB-SN~1998bw with an absolute magnitude of the SN at $\sim$23 days rest-frame of M$_{F140W}\approx-$19~mag. 
The SN is located in a host galaxy with complex morphology at a large ($\sim$6 kpc) offset in a region of relatively low surface brightness. 
The region around the SN has a modest star formation rate and is dominated by an intermediate mass-weighted age ($1.4\pm0.3$ Gyr) population, despite the apparent presence of a young, massive broad-lined Type Ic SN progenitor.  
These observations demonstrate that observations with {\em HST} and {\em JWST} can greatly extend the redshift range over which the GRB/FXT-SN connection can be studied, including in relatively low luminosity, X-ray rich events. They demonstrate little apparent evolution in the SN properties from local examples despite EP~240801a originating from an epoch 10~Gyr ago.

\end{abstract}


\section{Introduction}
The prompt emission in gamma-ray bursts (GRBs) can be broadly characterised by the duration and the hardness or fluence ratio, i.e.~the ratio between the spectral energies of lower and higher energy photons \citep{Barraud2003}. GRBs that are dominated by X-ray emission are called X-ray Flashes \citep[XRFs,][]{Heise2001, Kippen2001}, in contrast to classical and X-ray rich (XRR) GRBs with harder spectra that peak in gamma-rays \citep{Sakamoto2008}.
There was significant interest in the XRF population with their detection in instruments with relatively soft X-ray responses such as the {\em BeppoSAX}-WXM \citep{BeppoSAX} and the {\em HETE-2}/Fregate instruments. 
This population of XRFs has largely be rolled in with GRBs based on detections with the Neil Gehrels Swift Observatory \citep[{\em Swift,}][]{Gehrels2004}, with the majority of {\em Swift} bursts not classified on the GRB, XRR, XRF axis. However, with the advent of the Einstein Probe \cite[EP,][]{EP2015,Yuan2022} mission, there is renewed interest in the soft-end of the GRB-like population \citep[e.g.][]{lui25,levan25,oconnor25b}, with the wide-field X-ray telescope (EP/WXT) extending down to energies a factor of several lower than previous missions. This provides a new route to identify even softer outbursts than previously possible. 
Furthermore, there is significant interest in the relations between soft GRB-like objects and the population of Fast X-ray Transients (FXTs) that are bursts of soft X-ray photons ($\sim$0.3-10~keV), lasting from hundreds to thousands of seconds and have been identified in narrow-field focussing X-ray telescopes in the past two decades \citep[e.g.][]{Jonker_2013,AlpLarsson2020,qv22,qv23}. 

Thanks to EP, the sample size of FXTs and XRFs is growing rapidly, and includes rapidly available arcminute to arcsecond precision localisations, offering new opportunities to probe this soft end of the classical GRB distribution. 
In the GRB-related paradigm, XRFs represent the low photon energy end of the classical GRBs and might be explained by various intrinsic properties such as lower Lorentz factors in the GRB jets \citep[e.g.~dirty fireballs,][]{dermer1999}, but also by extrinsic effects such as the viewing angle between the observer and the jet \citep{Yamazaki2002}. Such a solution is supported because some XRFs are known to be associated to broad-lined supernovae (SNe) of Type Ic (Ic-BL SNe), with notable examples including
XRF~020903 \citep{Soderberg2005}, XRF~030723 \citep{Fynbo2004}
and GRB/XRF 060218 associated with SN2006aj \citep{Campana2006, Pian2006, Soderberg_2006}, strongly supporting a core-collapse origin in these events. However, in others SNe are ruled out to deep limits \citep{Soderberg2005,Levan2005}. 
A supernova interpretation is strengthened by the similarities between the environments of known XRFs and those of long GRB (LGRB) host galaxies \citep{Bi2018}, both of which show signs of high star forming rates and low-metallicities \cite[e.g.][]{Fruchter2006,Stanek2006}.
XRFs and LGRBs have therefore become one of the models that could explain the much more poorly understood populations of FXTs. FXTs have been identified for many years, but particular interest arises from the populations found in focussing X-ray telescopes which afforded accurate positions to be measured allowing to link them to a host galaxy \citep[e.g.,][]{Soderberg_2008, Jonker_2013}. Nevertheless, despite host galaxy associations, understanding their origin has been limited by the lack of contemporaneous follow-up observations, as almost all of them had been discovered years after the X-ray triggers. Various models have been proposed for their progenitors, including SN shock breakouts of core-collapse SNe \citep[e.g.,][]{Soderberg_2008, AlpLarsson2020}, binary neutron star mergers \citep[e.g.][]{Metzger2008} and the tidal disruption of a white dwarf by an intermediate-mass black hole \citep[e.g.,][]{Jonker_2013}.  
Now, over a hundred FXTs have been discovered by EP since its launch early 2024 and rapid reports of the events allow for follow-up observations. Several EP events are associated with GRBs \citep[e.g.,][]{levan25, Yin2024}, and the interpretation of many FXTs as similar physical events involving relativistic outflows from stellar scale engines has significant appeal, even in cases where there are deep limits on contemporaneous $\gamma$-ray emission \citep[e.g.,][]{2024GCN.38435....1R, 2025GCN.40703....1R, 2025GCN.39537....1R,1989BAAS...21..780H,
OConnor2025}. Indeed, several FXTs detected by EP without a detected gamma ray counterpart still show spectroscopic evidence of a Type Ic-BL SN \citep[e.g.,][]{vandalen2025, Sun2025, rastinejad2025}, providing a further plausible-link to the GRB progenitor population. 

The first FXT also classified as an XRF detected by EP was EP~240801a, detected with the Wide Field X-ray Telescope (WXT) on 1 August 2024 \citep{2024GCNEP240801a} (although the low peak energies inferred for several other events suggest they would also meet XRF criteria, but had no associated $\gamma$-ray data \citep[e.g.][]{Sun2025}. The Gamma-ray Burst Monitor (GBM) onboard the Fermi Space Telescope detected a weak contemporaneous signal \citep{Jiang2025}. 
\cite{Jiang2025} show that this is an extremely soft GRB with fluence ratio S(25~-~50~keV)/S(50~-~100~keV)~=~1.67$^{+0.74}_{-0.46}$ and hence classify it as an XRF. The redshift of EP~240801a was found to be $z \approx$~1.67 \citep{2024GCN.37013....1Q, Jiang2025}.
This values lies well beyond the most distant GRB/XRF SNe known \citep[e.g.~seen in][]{Finneran2025}, due to the observational limitations of the past 20-years of GRB follow-up.

In this letter, we present Hubble Space Telescope (HST) and James Webb Space Telescope ({\em JWST}) observations of EP~240801a, making use of the unique sensitivity and IR capabilities of {\em JWST} to reveal a Type Ic-BL SN and allowing to study the host galaxy. 

All magnitudes are given in the AB magnitude system. We assume a flat Lambda cold dark matter ($\Lambda$CDM) Planck cosmology \citep[\hspace{-5pt}][]{2020A&A...641A...6P} with H$_0$=67.7 km s$^{-1}$ Mpc $^{-1}$ and $\Omega_{m}$=0.31 throughout this work.

\section{Observations and data reduction}\label{sec:obs}

\begin{figure*}[ht]
    \centering
\gridline{\fig{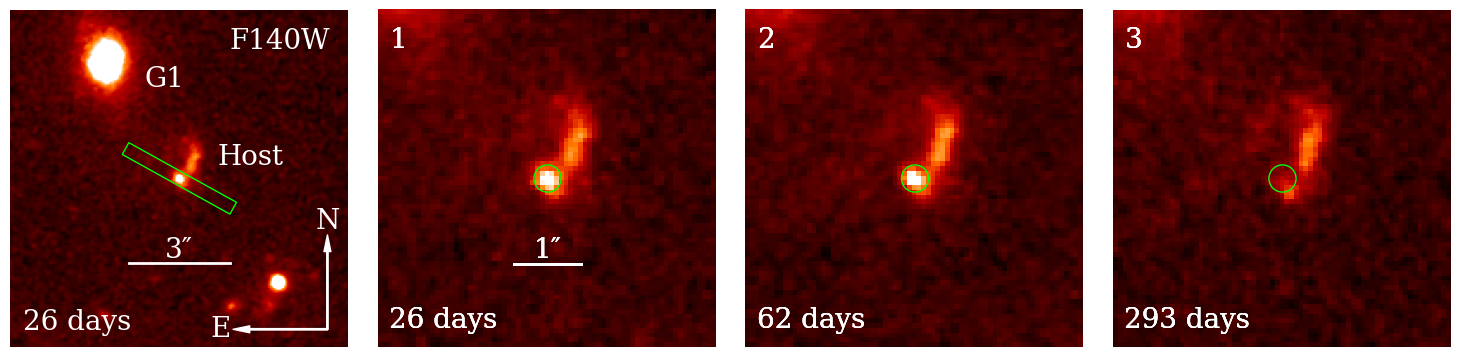}{\textwidth}{} } 
     \vspace{-20pt}
\gridline{\fig{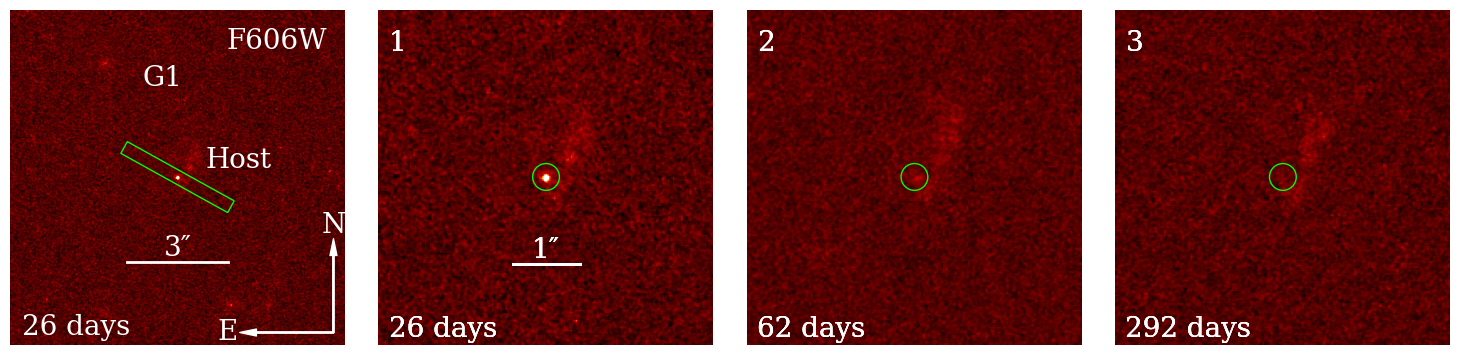}{\textwidth}{}}
  \vspace{-20pt}
\gridline{\fig{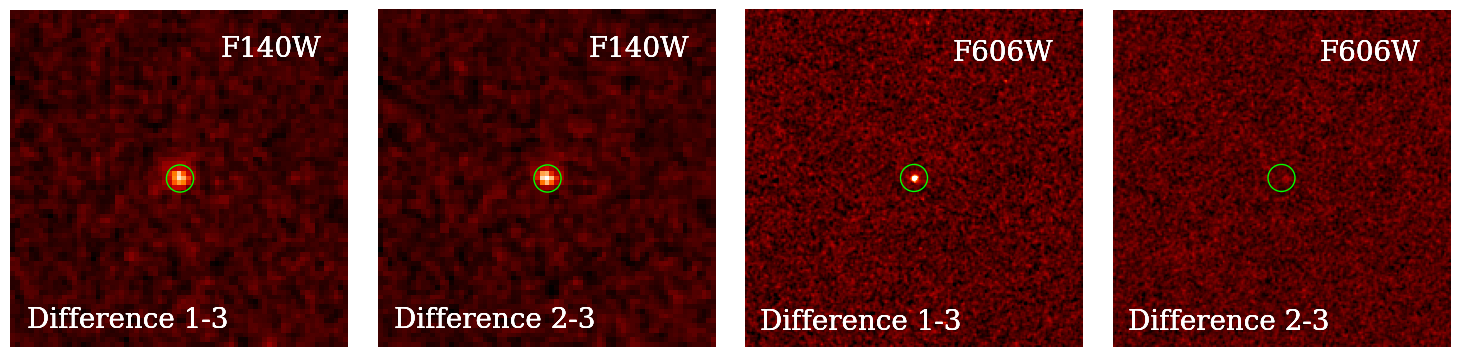}{\textwidth}{}}

     \vspace{-20pt}
        \caption{HST observations of the field of EP~240801a, showing all epochs as well as the resulting subtractions.  The labelled times are in observer-frame. The top row shows the F140W filter, while the second row shows the F606W filter. 
        In the left panels of the top two rows, the burst location can be seen to lie offset from the presumed host galaxy in a region of relatively low surface brightness, while another bright and extremely red galaxy (G1) can be seen proximate to the SN (see text). The location of the NIRSpec slit is marked with the green rectangles. 
        The other panels are a zoom in of the first. The third row shows all difference images on the scale of the zoomed images. The location of the SN is marked with a green circle with a radius of 0.2".}
        \label{fig:hst}
\end{figure*}

\subsection{Hubble Space Telescope (HST)}
We observed EP~240801a with HST at three epochs in the infra-red (IR) wide band filter F140W (which should be sensitive to supernova light) and ultraviolet-visible (UVIS) filter F606W (sensitive to the afterglow) of the Wide Field Camera 3 (WFC3). The data were taken under program IDs 17472, 17840; PI A. Levan and L. Izzo. The observations were taken approximately 25, 63 and 292 days after the WXT trigger (Table~\ref{tab:logHST}).
Images were aligned to sources in common to each frame, and subsequently drizzled to final pixel scales of 0.07 and 0.025 arcseconds per pixel for the IR and UVIS channels, respectively. The final epoch was then directly subtracted from the earlier observations to produce images containing only the transient light. The resulting images are shown in Figure~\ref{fig:hst}.

\subsection{James Webb Space Telescope (JWST)}
A single epoch of {\em JWST/NIRSpec} fixed slit spectroscopy was obtained of EP~240801a at 87.2 days after the WXT trigger (32.6 days in the rest-frame) under program 4569; PI A. Levan and L. Izzo. The spectrum was taken using the prism, providing wavelength coverage 0.6 to 5.3 $\mu $m.
Blind offsets were applied to place the 0.4 arcsecond (S400A1) slit on the burst location.

\section{results}
\subsection{Lightcurve}
Aperture photometry was performed using Source Extractor \citep[SE,][]{sextractor} in an 0.2" aperture measured on the difference images and corrected for encircled energy. All observations are listed in Table~\ref{tab:logHST} along with photometry in small apertures in the subtractions, and of the host galaxy as a whole. 

\begin{table*}
\movetableright=-1in
    \centering
    \caption{Observations of EP~240801a by {\em HST} and {\em JWST} discussed in this work. The photometry in the last column is the aperture photometry with 
    an 0.2" aperture and measured on the difference images, all values are corrected for encircled energy, but not corrected for Galactic extinction.}
    \resizebox{\textwidth}{!}{%
    \begin{tabular}{cccccccc}
Instrument	&	Epoch&	T since trigger &	T since trigger &	Exp. time 	&	Filter	& Apparent AB & Absolute AB \\
 &  (UT)	 & (observer-frame; d) & (rest-frame; d) & (s) & &Magnitude & \textcolor{blue}{Magnitude}  \\ \hline		
WFC3/IR	&	2024-08-26 11:04:51	&	25.58	&	9.57	&	3870	&	F140W	&	25.59 $\pm$ 0.03 & -18.86 $\pm$ 0.03\\		
WFC3/IR	&	2024-10-02 14:56:51	&	62.24	&	23.28	&	4423	&	F140W	&	25.64 $\pm$ 0.03 & -18.81 $\pm$ 0.03\\		
WFC3/IR	&	2025-05-21 08:34:52	&	293.48	&	109.79	&	4423	&	F140W	&	23.73 $\pm$ 0.02 $^{\dagger}$ & -20.72 $\pm$ 0.02$^{\dagger}$\\	
WFC3/UVIS	&	2024-08-26 12:39:05	&	25.15	&	9.41	&	2132	&	F606W	&	26.47 $\pm$ 0.07 & -17.98 $\pm$ 0.07\\		
WFC3/UVIS	&	2024-10-02 11:26:56	&	62.60	&	23.42	&	4000	&	F606W	&	28.60 $\pm$ 0.28 & -15.85 $\pm$ 0.28\\	
WFC3/UVIS	&	2025-05-19 23:33:04	&	291.60	&	109.09	&	4440	&	F606W	&	24.89 $\pm$ 0.28 $^{\dagger}$ & -19.56 $\pm$ 0.28$^{\dagger}$\\

       JWST/NIRSpec & 2024-10-27 15:37:49 & 87.19 & 32.62 & 3501 & PRISM/CLEAR & - & -\\
         \hline
    \end{tabular}%
    }
    \label{tab:logHST}
The photometric values marked with a $\dagger$ are the magnitudes of the entire underlying host galaxy in a 0.9\arcsec~ aperture.
\end{table*}

\begin{figure*}
    \centering
    \includegraphics[width=0.8\linewidth]{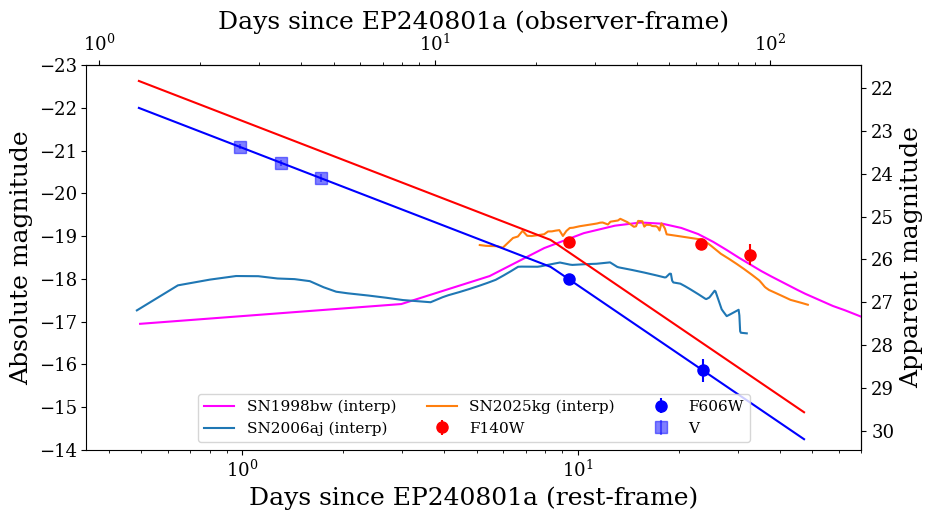}
    \caption{The lightcurve of EP~240801a in absolute magnitude (left axis) and apparent magnitude (right axis) versus the rest-frame (bottom axis) and observer-frame (top axis) time since the EP trigger in days. The early-time V-band observations from \cite{Jiang2025} are shown with square markers, while the HST F606W and F140W observations are shown with dots in blue and red, respectively. The last point in the F140W band is taken from the JWST/NIRSpec spectrum. A broken power-law fit of the afterglow in the V and F606W band observations is shown with a blue line. The red line is the same fit shifted towards the F140W filter, using the spectral index of $\beta=0.69$.
    The HST~F140W magnitudes extracted from the interpolated lightcurves of SN~1998bw, SN~2006aj and SN~2025kg are shown with solid magenta, orange and light blue lines, respectively, for comparison.
    }
    \label{fig:lightcurve}
\end{figure*}

The F606W (broad V-R) and F140W (broad H) lightcurve is shown in Figure~\ref{fig:lightcurve}.
Comparing the early-time \citep{Jiang2025} V-band observations with the late-time HST F606W observations, we see a jet break in the afterglow V-band lightcurve. A broken power-law fit through the early V-band observations has temporal index $\alpha_1=0.86$ ($F \propto t^{-\alpha}$), while the two HST epochs of F606W can be fitted with a power-law with $\alpha_2=2.1$, suggestive of substantial steepening with a break at $t_b =8.3$ days rest-frame. This is most likely due to the jet-break and consistent with the inferred break-time of $\sim$70~ks from \cite{Jiang2025}. The flat lightcurve seen in the F140W observations is evidence for the associated supernova (Section~\ref{sec:sn}).

\subsection{NIRSpec spectrum}
The trace in the reduced two dimensional spectrum (see Figure~\ref{fig:NIRSpec}) is clearly extended, showing light from both the host galaxy and the optical transient. 
To separate the counterpart and host galaxy light we extract the spectrum on a row-by-row basis using the {\em JWST} pipeline \citep{Bushouse2023zjwstpipeline} one dimensional extraction tool {\tt extract1d}. 

The spatially resolved spectra for the different rows of pixels in the NIRSpec detector are shown in the middle panel of Figure~\ref{fig:NIRSpec}, and clearly shows broad supernova like features in addition to emission lines from an extended host galaxy. A simple thermal fit via \texttt{Redback} \citep{Sarin2024redback} yields a poor reduced chi-squared because of the 
broad features, but gives a
temperature of $T_{BB}\approx6000$~K and a radius of $\sim 2\times10^{15}$ cm. The spectrum is discussed further in Section~\ref{sec:sn}.

\subsection{Host galaxy}
The host galaxy of EP~240801a is clearly visible in our HST imaging, and has a complex morphology. EP~240801a is offset approximately 0.3 arcseconds (2.6~kpc in projection) to the South-West of a bright knot of emission with emission lines (H$\alpha$, H$\beta$, [OII], [OIII] and HeI) at the same redshift as EP~240801a. This knot of emission appears to be part of a larger galaxy whose centroid location is $0.63 \pm 0.06$ arcseconds (6 kpc in projection) from the transient (see Fig.~\ref{fig:hst}). The host has magnitudes of F606W=24.89$\pm$0.28~mag and F140W=23.73$\pm$0.02~mag measured in an 0.9\arcsec~aperture. The half light radius, as measured via SE is 0.59\arcsec.

\section{Discussion}

\subsection{Supernova discovery}
\label{sec:sn} 

Between our first two HST epochs (25 and 62 days post burst) the counterpart fades by $2.13 \pm 0.29$ mag in F606W, but only by $0.05 \pm 0.04$ mag in F140W. 
This provides strong photometric evidence for an SN, as line blanketing in the UV would at $z=1.67$ make an SN very faint in the F606W filter (rest-frame $\sim 2250$\AA), while the IR F140W light would be sensitive to the SN spectral peak (rest-frame $\sim 5250$\AA). Assuming that the F606W is purely afterglow with a spectral shape of $\beta = 0.69 \pm 0.02$ \citep{Jiang2025}, we can extrapolate the afterglow flux in F140W (see Appendix~\ref{Appendix:photometry}). Then, using the first epoch F140W magnitude, we can also calculate the SN flux. We find that the first epoch F140W observation contained substantial contributions from both SN and afterglow light ($0.17 \pm 0.02$ $\mu$Jy from the afterglow and 0.05 $\pm$ 0.02 $\mu$Jy from the SN, magnitudes in Table.~\ref{tab:all_magnitudes}), although any small SN contribution in F606W, or systematics in the afterglow model could impact this. By the time of the second {\em HST} observation and the JWST spectrum the source should be entirely dominated ($>90\%$) by the SN light.
The absolute magnitude of the SN, \sn, is $M_{F140W}=-18.82\pm~0.02$~mag at $\sim$23 days in rest-frame (the time of our second epoch observations), consistent with the values found by e.g.~\cite{Taddia2019} for Ic-BL SNe.

We compare the event to SN~1998bw, a prototypical SN related to a GRB \citep{Clocchiatti2011}, to SN~2006aj, which is related to an XRF \citep{Mirabal_2006} and to SN~2025kg, an FXT with a Type Ic-BL SN, but no gamma-rays \citep{rastinejad2025,Eyles-Ferris2025}.
We use the known redshifts of these events to obtain the rest-frame filter observations of the lightcurves and use these to interpolate to a lightcurve in the F140W filter. The absolute magnitude of SN~2025kg at 23 days and SN~1998bw at both 9 and 23 days rest-frame is similar to that of \sn, suggesting a similar luminosity and evolution. \sn~is brighter than SN~2006aj, but well within the expected range for GRB-SNe \citep{Hjorth_2013}. We do note that since there is a substantial afterglow contribution to the first epoch, a rather different SN lightcurve morphology is required, which rises to a later, but narrower peak than seen in SN~1998bw.

\begin{figure*}[h!]
    \centering
    \includegraphics[width=\linewidth]{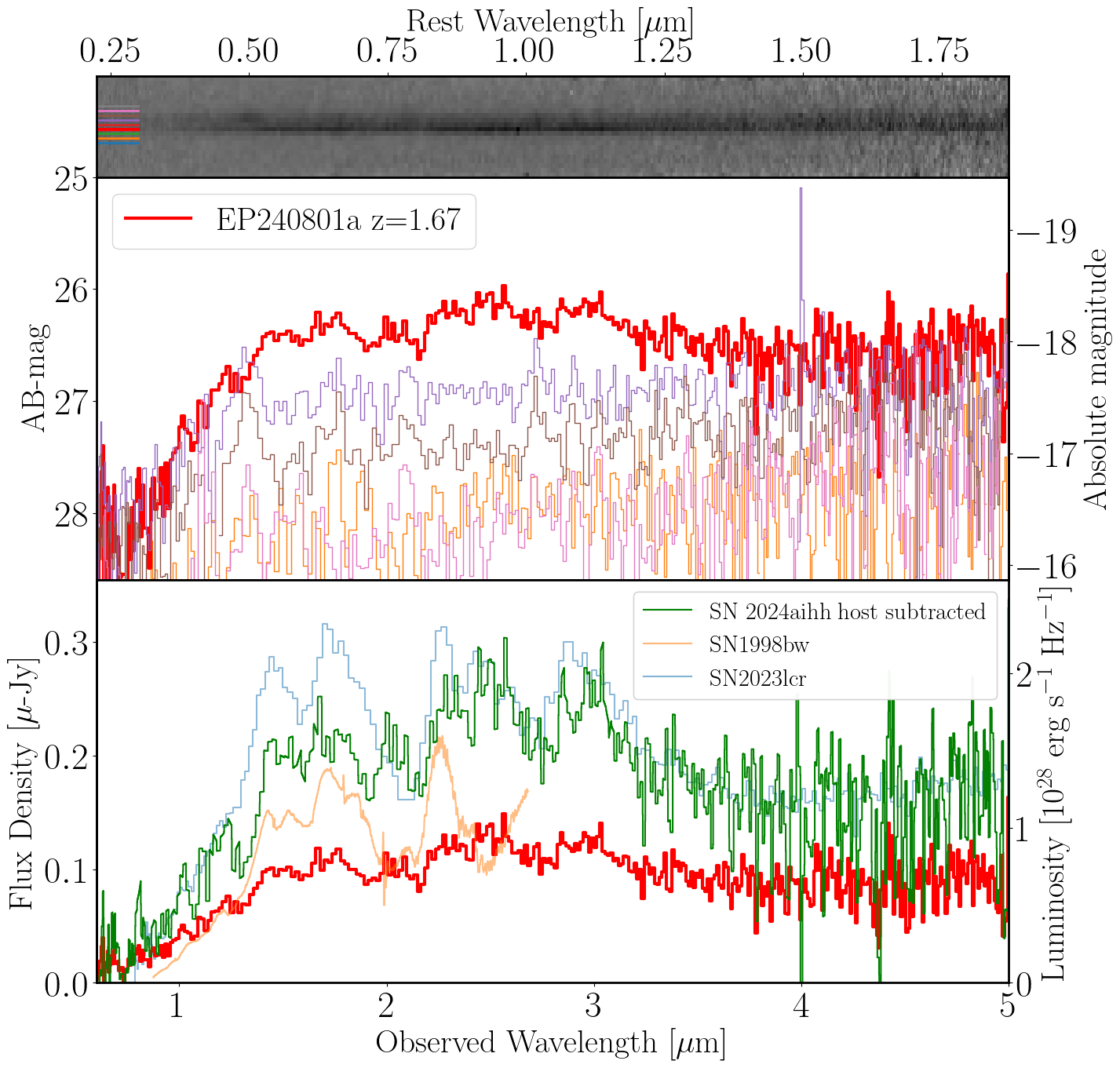}
    \caption{
    \textit{Top panel:} The 2D {\em JWST/NIRSpec} spectrum in rest wavelength (top axis) and observed wavelength (bottom axis) in $\mu$m.
    \textit{Middle panel:} Single pixel extractions as indicated with the small horizontal coloured lines in the 2D spectrum. The spectrum containing \sn~is shown in red. The left axis shows the apparent magnitude and the right axis the absolute magnitude.
    \textit{Bottom panel:} The extraction of \sn~(bright red) compared to a {\em JWST/NIRSpec} spectrum of SN~2023lcr (light blue) at 27 days rest-frame and a spectrum of SN~1998bw (orange) at 33 days rest-frame. On the left vertical axis the flux density in $\mu$Jy is shown and on the right axis the luminosity density in 10$^{28}$ erg s$^{-1}$ Hz$^{-1}$. 
    The host-subtracted and aperture-corrected spectrum of \sn~is shown in green. The spectrum generally matches well to the overall shapes of the templates.
    }
    \label{fig:NIRSpec}
\end{figure*}

Even stronger evidence for the associated supernova arises from the NIRSpec spectroscopy. In particular, this shows a series of broad spectral features and a strong drop-off in the rest-frame UV that strongly resemble those seen in other Type Ic-BL SNe and those observed in other LGRBs, leading to the classification of the supernova \sn. In the bottom panel of Figure~\ref{fig:NIRSpec} we show a comparison between the {\em JWST} spectrum of \sn~and spectra of two well known GRB-SNe examples, this includes the proto-type SN~1998bw at $z=0.0085$ at a similar spectral epoch to our observations ($\sim$33 days rest-frame), as well as a {\em JWST} spectrum of SN~2023lcr \citep{MartinCarilloprep} consistent with an on-axis GRB at $z=1.03$ at $\sim$27 days rest-frame. The latter has been observed with NIRSpec and so covers close to the entire spectral range of our observations. We note that the strength of the discrete spectral features in \sn~is weaker than in the templates, although the signal-to-noise ratio is also lower and some host contribution is likely. Although the resolution of the NIRSpec prism is low, the broad Type Ic-BL SN features are still well resolved. The association of EP~240801a with a type Ic-BL SN further strengthens the connection between XRFs and SNe.

Since \sn~is embedded within its host galaxy it is likely that this contributes to the light. This host contamination cannot be directly removed without a late time host-only observation. However, we attempt to do so using the spatially resolved information in the NIRSPEC spectrum in concert with the available late-time HST imaging. 

We first measure the flux in the third epoch HST~F140W image in a rectangular aperture the size of a single pixel row in {\em JWST/NIRSpec} (0.4\arcsec $\times$ 0.1\arcsec) on the location of the {\em JWST/NIRSpec} position where we extracted the SN spectrum. We find a magnitude of $m \approx 28.1$~mag, although note that photon noise and the point spread function (PSF) differences between HST and JWST impact this and cannot readily be calculated for a complex extended source, providing for some additional photometric error. 

We then extract a ``host galaxy" spectrum by averaging the NIRSPEC spectrum in two rows 0.2\arcsec~ away from the transient position, indicated with the purple and orange lines in Fig.\ref{fig:NIRSpec}, and scale this to the F140W magnitude measured in the HST imaging. We subsequently subtract this spectrum from the row containing the transient to create a host subtracted spectrum. 

Finally, since we only use a single row extraction, we also need to correct for the light loss due to the PSF of a point source being larger than one pixel.
We find this wavelength dependent aperture correction by obtaining the flat-fielded and dark-subtracted NIRSpec fixed slit observation of a standard star from the MAST archive (target J1743045 observed under program 1128; P.I.~N.~Luetzgendorf) and re-extract the spectrum on the central pixel with a single pixel width following the same procedure as done for our data. We calculate the ratio between this single row extraction and the default extraction to find the aperture correction that we apply to our spectrum. Because the NIRSPEC pixels are comparable in size to the PSF the precise aperture correction will depend on the source location within the pixel, and this induces some additional errors. However, measuring the residual location in our HST images suggests that our source was close to ideally placed, and so the ultimate correction is likely similar to that of the standard. 

The resulting SN spectrum is shown with the green line in Figure~\ref{fig:NIRSpec}. It matches the spectral shape of SN~2023lcr after $\sim$2$\mu$m observer-frame well, but \sn~is weaker towards the bluer end, which could be due to a lower signal-to-noise ratio. Indeed \sn~at $z\sim1.67$ looks rather similar to the GRB-SNe at $z\sim0$, despite it being 10~Gyr ago.

As noted above the strength of the line features in the \sn~spectrum appears to be lower than in the template spectra. We suggest this is likely due to a combination of lower signal-to-noise ratio and host contamination. To test if this is a plausible solution we adopt the opposite test of ``adding" the host light to the template spectra of SN~1998bw and AT2023lcr as shown in Figure~\ref{fig:NIRSpec_hostadded}. 
For this, we use the host galaxy spectrum as described before and combine 62\% of it with 38\% of the SN spectra\onecolfootnote{These percentages are determined by calculating the ratio between the mean flux of the ``host galaxy" spectrum and that of the single pixel extraction at the SN location.}.
This demonstrates that indeed the host contribution can dilute the strength of the broad features, and that there is some degeneracy between the contribution of the SN and host in the spectrum alone. However, our HST photometric measurements do provide a clean measurement of the host for subtraction in the lightcurve. We further note that in our host subtracted version of the NIRSpec spectrum, the resulting photometry using \texttt{pyphot}, provides F140W=$25.89\pm0.25$~mag, consistent with the lightcurve of SN~1998bw at comparable epochs, as seen from the last point in Figure~\ref{fig:lightcurve}. The uncertainty on this magnitude is estimated to include all the uncertainties on the host spectrum as describes before.

\subsection{Host galaxy}

\subsubsection{Local environment}
To assess the local environment of the EP~240801a progenitor, we calculate the fractional flux following \cite{Fruchter2006}, which is particularly useful for irregular galaxies like the host of EP240801. We obtain the transient position in our subtracted images, which pinpoints the source with an uncertainty $<<1$ HST pixel. 
With SE, we extract the host galaxy pixels using a detection threshold of 5$\sigma$ on the third F140W epoch and use the resulting segmentation map (Fig.~ \ref{fig:segmentationmaps}).
The resulting fractional flux is $\sim$0.4, which is relatively low, but not unheard of for GRB hosts.
The host normalised offset of EP~240801a is $R/R_h\approx$~1.1, where $R_h$ is the half-light radius of the galaxy, also suggestive of a burst in a somewhat (but not exceptionally) unusual location.

It has previously been suggested that even when bursts lie at large Galactocentric radii they are still found preferentially in regions of relatively high surface brightness \citep{Blanchard2016} (i.e. a preference to be found on star-forming knots of emission rather than far out in faint regions of the host). Indeed, if we calculate our fractional flux while masking a circular region with r~\textless~0.5~$\times R_h$ centred on the centre of the galaxy \citep[as in][]{Lyman2017} as shown in the right panel of Figure \ref{fig:segmentationmaps}, we then find a fractional flux of $\sim$0.6, showing that indeed this SN appeared on one of the brighter pixels of the outer region of the galaxy.

\subsubsection{SED fitting}
We use the Bayesian Analysis of Galaxies for Physical Inference and Parameter EStimation \citep[BAGPIPES;][]{bagpipes} tool to fit a spectral energy distribution (SED) to the extracted JWST/NIRSpec host galaxy spectrum, to infer the properties of the the region of the underlying galaxy next to the transient at $\sim$2~kpc away (the pixels indicated with the purple, brown and pink coloured lines in the top panel of Fig.~\ref{fig:NIRSpec}). These pixels have been chosen because they are not contaminated by transient light while they still allow for a significant detection of the galaxy light. Our best fit is shown in Fig.~\ref{fig:bagpipes_JWST_host}. This fit indicates a modest star formation rate, and interestingly implies that the galaxy is dominated by a stellar population of intermediate age $1.4\pm0.3$~Gyr. This is consistent with the relatively weak emission lines observed, as can be seen in Fig.~\ref{fig:bagpipes_JWST_host}.  
The metallicity of the host is $0.57\pm0.06$ Z$_{\odot}$ from the SED fit, towards the higher end of metallicities observed in GRB host galaxies \citep{palmerio2019}, although we note this is not a measurement of the metallicity directly under the transient location, but about 2~kpc away. We also note that we cannot directly compare this metallicity to those for GRB hosts obtained through line-ratios, due to the low resolution of the NIRSpec spectrum leaving us unable to separate individual lines.
The host galaxy has a low A$_V$=0.03$^{+0.03}_{-0.02}$ according to our BAGPIPES fit, using the \citet{Calzetti2000} dust model.

We do not have a full spectrum covering the entire host galaxy, and so to estimate the global properties (e.g.~SFR, stellar mass) we assume that the properties measured in the slit are consistent with those of the galaxy as a whole and scale them according to the fraction of light in the slit in the HST~F140W band compared to the F140W magnitude of the whole galaxy. In particular, we fold the measured NIRSpec spectrum through the F140W filter response using \texttt{pyphot} \citep{zenodopyphot}, obtaining a magnitude of F140W~=~26.1~$\pm$~0.1~mag, compared to the integrated magnitude of the host galaxy of F140W=23.7 $\pm$ 0.1~mag. A direct scaling then implies SFR=3.9$\pm0.4$ M$_{\odot}$ yr$^{-1}$, and a stellar mass of M$_*$=(0.4$\pm$0.1)$\times$10$^9$~M$_{\odot}$. 

\subsection{Galaxy G1}
We also note the presence of a very red galaxy $\sim$4.3 arcseconds in projection to the North-East of EP~240801a (marked G1 in Figure~\ref{fig:hst}). This galaxy is an Extremely Red Object (ERO) also detected in the legacy survey and with WISE. 

In a stack of all the HST~F140W images there is evidence that the light of this galaxy extends towards the transient position, as well as possible tidal features that could indicate that the host of EP~240801a is currently involved in an ongoing interaction with this galaxy. Although it is also possible that this represents a chance alignment, a tidal interaction would readily explain both the irregular morphology of the host galaxy and the presence of star formation at the burst location. A BAGPIPES fit to the photometry (given in  Table~\ref{tab:G1_magnitudes}), provides a photometric redshift of $z \approx$~1.3~$\pm$0.1 (the best fit is shown in Figure \ref{fig:bagpipes_ERO}), which is marginally inconsistent with the redshift of EP~240801a, although ultimately a spectroscopic redshift will be necessary to ascertain any connection.

\section{Conclusion}
In this Letter, we have presented {\em HST} and {\em JWST} observations of the field of EP~240801a. We find evidence of a Type Ic-BL SN, \sn, similar to SN~1998bw and SN~2023lcr. This makes EP~240801a another XRF associated with such an SN.
The apparent SN magnitude at $\sim$23 days rest-frame is $m_{F140W}=25.64\pm0.03$~mag, too faint and red to be detected by ground-based telescopes. Its absolute magnitude is M$_{F140W}=-18.82\pm$0.02~mag.

The NIRSpec spectrum of \sn~exhibits relatively weak features compared to the features in the spectra of SN~1998bw and SN~2023lcr. This is likely a combination of a lower signal-to-noise ratio and some host contamination in EP~240801a that is not present in archival spectra of SN~2023lcr and SN~1998bw. After host-subtraction and aperture correction, the SN resembles the comparison GRB-SNe at $z\sim0$, even though EP~240801a is 10 Gyr ago.

The SN is offset by $\sim6$ kpc (projected) from the host galaxy's centroid location, and in the vicinity ($\sim2.6$ kpc) of a region of bright emission. The fractional light of the pixel containing the SN is $\sim$0.4.
Stellar population fitting of the local environment around the SN yields an intermediate age of $1.4\pm0.3$ Gyrs and a relatively high metallicity of $0.57 \pm 0.06$ Z$_{\odot}$.

These observations demonstrate the promise of HST and JWST to study the connection between GRBs, XRFs and FXTs at much higher redshift than previously possible, opening the possibility of studying any possible evolution in the population progenitors across cosmic time. Indeed, this observation, in concert with some photometry at much higher redshift \citep{levan25c} suggests that in this small sample there is little evolution in the properties of GRB-supernova across 13 billion years of cosmic time.

\section*{Acknowledgements} 

A.P.C.H., P.G.J., J.N.D.D., J.Q.V., and J.S.S.~ are supported by the European Union (ERC, Starstruck, 101095973, PI Jonker). Views and opinions expressed are however those of the author(s) only and do not necessarily reflect those of the European Union or the European Research Council Executive Agency. Neither the European Union nor the granting authority can be held responsible for them.
J.Q.V. additionally acknowledges support by the IAU-Gruber foundation.

LI acknowledges financial support from the INAF Data Grant Program `YES' (PI: Izzo) {\it Multi-wavelength and multi messenger analysis of relativistic supernovae}.

This work was supported by a research grant (VIL54489) from VILLUM FONDEN.

This work is based in part on observations made with the NASA/ESA/CSA James Webb Space Telescope. The data were obtained from the Mikulski Archive for Space Telescopes at the Space Telescope Science Institute, which is operated by the Association of Universities for Research in Astronomy, Inc., under NASA contract NAS 5-03127 for JWST. These observations are associated with program \#4569. The observation analysed in this work can be accessed via \dataset[DOI: 10.17909/2s7b-1197]{https://doi.org/10.17909/2s7b-1197}. 

This research is based in part on observations made with the NASA/ESA Hubble Space Telescope obtained from the Space Telescope Science Institute, which is operated by the Association of Universities for Research in Astronomy, Inc., under NASA contract NAS 5–26555. These observations are associated with programs 17474 and 17840 and can be accessed via \dataset[DOI: 10.17909/2jt9-qk86]{https://doi.org/10.17909/2jt9-qk86}.

Data for this paper has in part been obtained under the International Time Programme of the CCI (International Scientific Committee of the Observatorios de Canarias of the IAC) under programm ID ITP24 PI Jonker with the GTC operated on the island of La Palma by the Roque de los Muchachos.

\bibliography{references}{}
\bibliographystyle{aasjournalv7}

\appendix
\section{Host galaxy contamination}

\begin{figure}[h]
    \centering
    \includegraphics[width=0.75\linewidth]{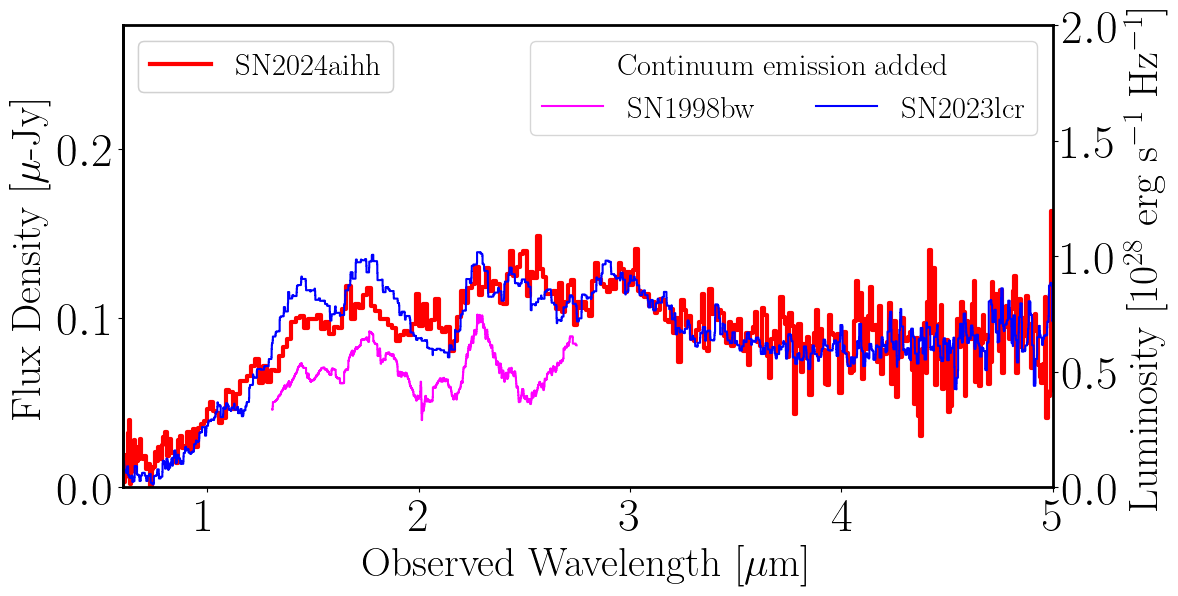}
    \caption{The extraction of \sn~(bright red) compared to a {\em JWST/NIRSpec} spectrum of SN~2023lcr (blue) at 27 days rest-frame and a spectrum of SN~1998bw (magenta) at 33 days rest-frame, both models are modified by 62\% underlying continuum emission.
    }
    \label{fig:NIRSpec_hostadded}
\end{figure}

\section{Segmentation map host galaxy}

\begin{figure}[h]
    \centering
    \includegraphics[width=0.75\linewidth]{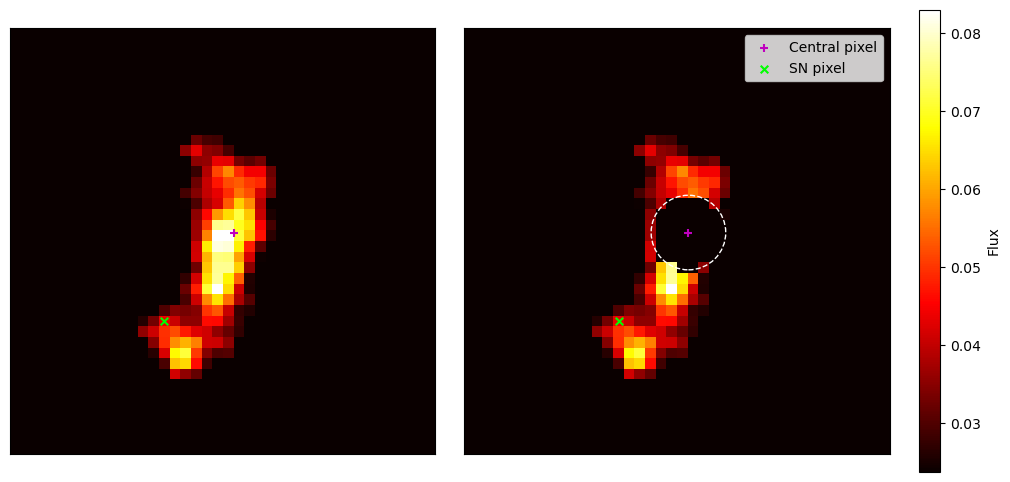}
    \caption{The segmentation map of the host galaxy of EP~240801a obtained with SE using a detection threshold of 5$\sigma$. The left panel shows the segmentation map of full galaxy, while the right panel shows the segmentation map with the inner region (r \textless 0.5$\times R_h$) excluded. This inner region is shown with the white dashed circle and the central pixel is marked with a magenta ``+"-sign. The pixel containing the SN is marked with a lime-green cross. The colours of the map indicate the relative brightness of the pixels.}
    \label{fig:segmentationmaps}
\end{figure}

\section{Galaxy fitting}
\begin{figure}[ht]
    \centering
    \includegraphics[width=0.8\linewidth]{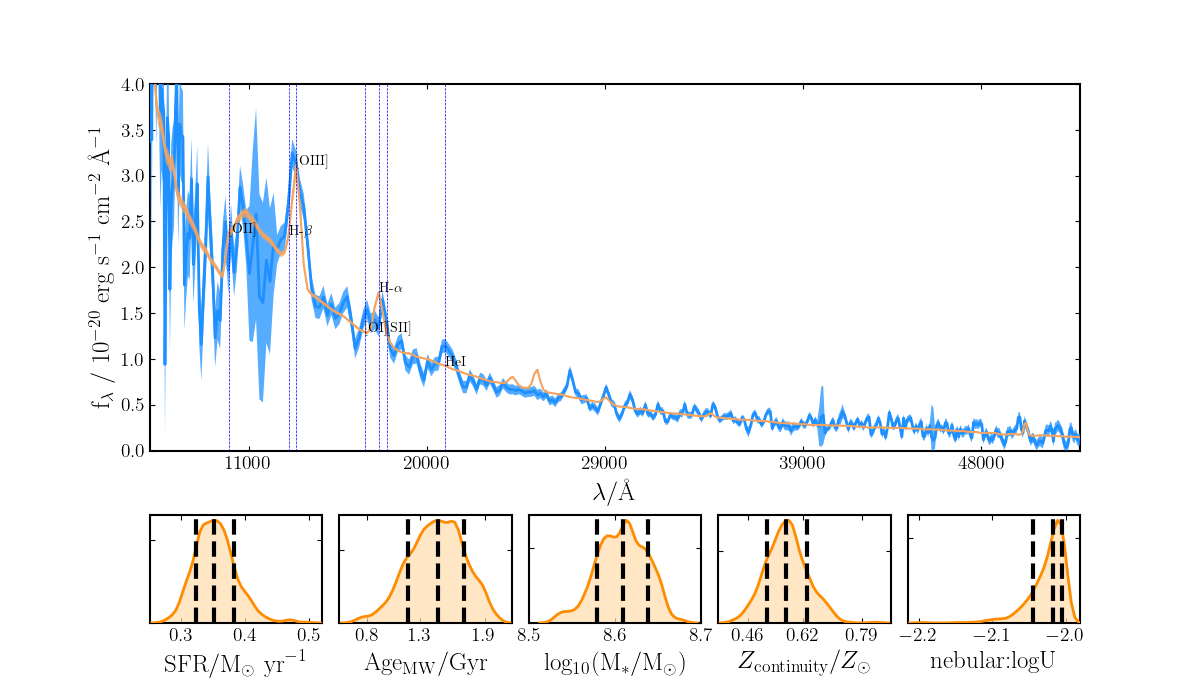}
    \caption{\textit{Top panel:} The host galaxy spectrum shown in blue of the region around EP~240801a extracted from three pixel rows in the {\em JWST/NIRSpec} spectrum in F$_{\lambda}$ in erg s$^{-1}$ cm$^{-2}$ \AA$^{-1}$ versus observed wavelength in \AA. The shaded region shows the 1$\sigma$ uncertainty. The best-fit obtained with BAGPIPES is shown in orange. 
    \textit{Bottom panels:} The posterior distributions for five fitted parameters are shown, from left to right: SFR, age, galaxy stellar mass, metallicity and the nebular ionization parameter. The 16th, 50th and 84th percentile posterior values are indicated by the vertical dashed black lines in each subplot. }
    \label{fig:bagpipes_JWST_host}
\end{figure}

The extracted {\em JWST/NIRSpec} host galaxy spectrum was fitted using BAGPIPES to infer the host galaxy parameters (Figure \ref{fig:bagpipes_JWST_host}). The \cite{leja2019} continuity non-parametric star formation history model was considered with Student's t-distribution priors (-10, 10) for the star formation in each age bin. The edges of the eight age bins are logarithmically spaced according to the age of the universe at $z=1.67$. Additional free parameters are the formed mass, metallicity, \cite{Calzetti2000} dust parameter $A_V$, the nebular ionization parameter $U$ and a small region around the known redshift of the galaxy (1.66, 1.68).

\begin{figure}[h]
    \centering
    \includegraphics[width=0.8\linewidth]{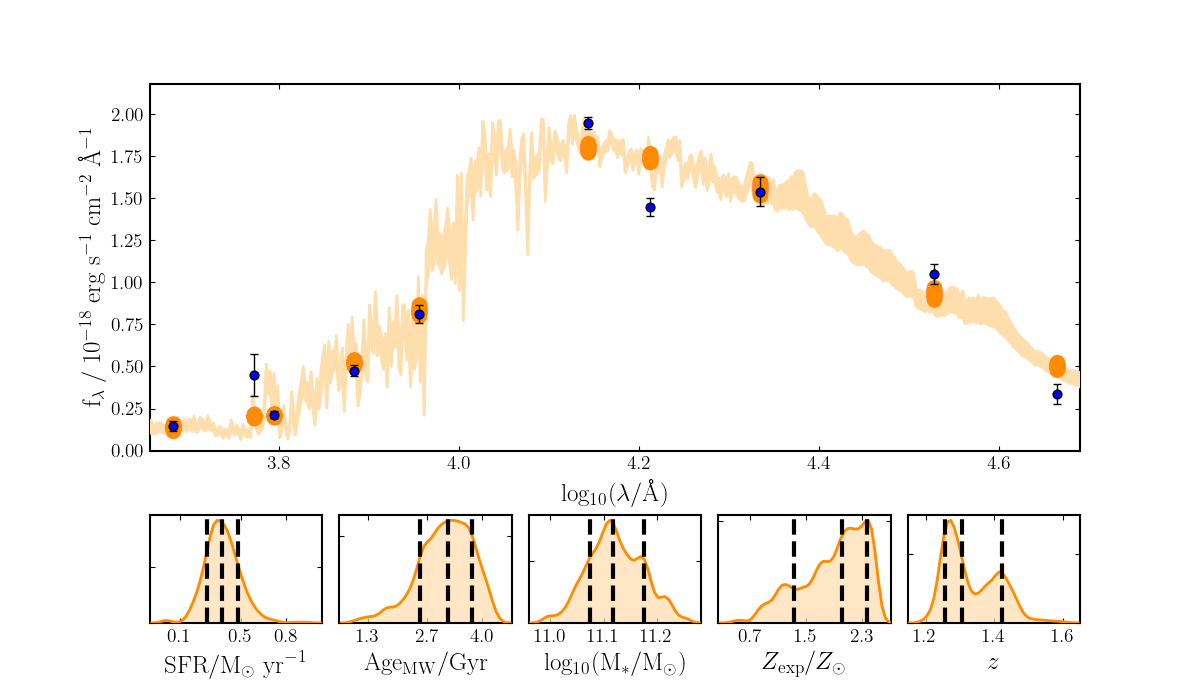}
    \caption{\textit{Top panel:} The ERO galaxy photometry in HiPERCAM \textit{griz}, HST F140W and F606W, EMIR H and Ks and WISE W1 and W2 shown in blue F$_{\lambda}$ in erg s$^{-1}$ cm$^{-2}$ \AA $^{-1}$ versus observed wavelength in \AA. The best-fit obtained with BAGPIPES is shown in orange with the model photometry with darker markers on top. 
    \textit{Bottom panels:} The posterior distributions for five fitted parameters are shown, from left to right: SFR, age, galaxy stellar mass, metallicity and the redshift. The 16th, 50th and 84th percentile posterior values are indicated by the vertical dashed black lines in each subplot.}
    \label{fig:bagpipes_ERO}
\end{figure}

The ERO galaxy photometry is provided in Table~\ref{tab:G1_magnitudes} and was fitted using BAGPIPES to extract the galaxy parameters and the photometric redshift (Figure \ref{fig:bagpipes_ERO}). We used an exponentially declining star formation history model. Additional free parameters are the formed mass, metallicity, and \cite{Calzetti2000} dust parameter $A_V$.

\begin{table}[h]
    \centering
        \caption{AB magnitudes of G1 in different filters taken with multiple telescopes/instruments. Values are not corrected for Galactic extinction.}
    \begin{tabular}{c|c|c}
       Telescope/Instrument  & Filter & AB Magnitude \\ \hline 
         GTC/HiPERCAM & g &  26.28 $\pm$ 0.22 \\
         GTC/HiPERCAM & r &  25.30 $\pm$ 0.10 \\
         GTC/HiPERCAM & i &  23.98 $\pm$ 0.07 \\
         GTC/HiPERCAM & z &  23.04 $\pm$ 0.07 \\
         HST/WFC3 & F140W & 21.15 $\pm$ 0.02 \\
         HST/WFC3 & F606W & 24.60 $\pm$ 0.30 \\
         GTC/EMIR & H & 21.08 $\pm$ 0.04 \\
         GTC/EMIR & Ks & 20.45 $\pm$ 0.06 \\
         WISE & W1 & 19.90 $\pm$ 0.06 \\
         WISE & W2 & 20.45 $\pm$ 0.19 \\
         \hline 
    \end{tabular}
    \label{tab:G1_magnitudes}
\end{table}

\newpage
\section{Photometric measurements}
\label{Appendix:photometry}
As we have presented multiple photometric measurements for the transient and different regions within the host galaxy throughout the paper, we provide Table~\ref{tab:all_magnitudes} listing all the magnitudes obtained through different methods. We used \texttt{pyphot} to measure the photometric values from the different pixel extractions of the NIRSPEC spectrum. Where the afterglow and SN light are separated for F140W, the afterglow flux is calculated from the spectral index of the afterglow $\beta=0.69$ \citep{Jiang2025} using ${\rm flux}_{F606W}*(\frac{\lambda_{F140W}}{\lambda_{F606W}})^{\beta}$. The SN light is then the rest of the flux measured in the 0.2\arcsec~aperture.

\begin{table}[]
    \centering
        \caption{AB magnitudes for the transient and its environment used throughout this work. The magnitudes have been obtained with different methods, as listen in the second column. Magnitudes obtained from the NIRSpec spectrum are derived using \texttt{pyphot}.}
    \begin{tabular}{l|c|c|c}
       Target & Method & Filter & Magnitude\\ \hline
       \multicolumn{4}{c}{\textit{Transient}} \\ \hline  
    Afterglow 9.41 rest-frame days & 0.2\arcsec~ aperture & F606W & 26.47$\pm$0.07 \\
    Transient 9.57 rest-frame days & 0.2\arcsec~ aperture & F140W & 25.59$\pm$0.03\\
    Afterglow 9.57 rest-frame days & Extrapolated & F140W & 25.84$\pm$0.03\\
     & from F606W &  & \\
    SN 9.57 rest-frame days & Transient -& F140W & 27.31$\pm$0.03\\
     & afterglow &  & \\
    Afterglow 23.42 rest-frame days & 0.2\arcsec~ aperture & F606W & 28.60$\pm$0.28\\
    SN 23.28 rest-frame days & 0.2\arcsec~ aperture & F140W & 25.64$\pm$0.03 \\
    SN 32.62 rest-frame days & NIRSpec spectrum & F140W & 25.89$\pm$0.25 \\
    \hline
       \multicolumn{4}{c}{\textit{Host galaxy}} \\ 
         \hline 
Full host galaxy & 0.9\arcsec~ aperture & F140W & 23.73$\pm$0.02 \\
Full host galaxy & 0.9\arcsec~ aperture & F606W &  24.89$\pm$0.28 \\
Under the SN & 0.4\arcsec~ $\times$ 0.1 \arcsec~ aperture & F140W & $\sim$28.1 \\  
Next (0.2\arcsec-0.4\arcsec) to SN & NIRSpec spectrum & F140W &  26.1$\pm$0.1 \\ \hline
    \end{tabular}
    \label{tab:all_magnitudes}
\end{table}

\end{document}